\documentclass{elsart}
\usepackage{graphicx}
\usepackage{amssymb}

\begin{document}

\begin{frontmatter}

\title{Avalanche dynamics of an idealized neuron function in the brain on uncorrelated random scale-free network}

\author[jwl]{Kyoung Eun Lee}
\author[jwl]{Jae Woo Lee\corauthref{cor1}}
\ead{jaewlee@inha.ac.kr}

\corauth[cor1]{Tel.: 82+32+8607660; Fax: 82+32+8727562}
\address[jwl]{Department of Physics, Inha University, Incheon 402-751, Korea}

\begin{abstract}
 We study a simple model for a neuron function in a collective brain
system. The neural network is composed of uncorrelated random
scale-free network for eliminating the degree correlation of
dynamical processes. The interaction of neurons is supposed to be
isotropic and idealized. This neuron dynamics is similar to
biological evolution in extremal dynamics with isotropic locally
interaction but has different time scale. The evolution of neuron
spike takes place according to punctuated patterns similar to the
avalanche dynamics. We find that the evolutionary dynamics of this
neuron function exhibit self-organized criticality which shows
power-law behavior of the avalanche sizes. For a given network, the
avalanche dynamic behavior is not changed with different degree
exponents of networks, $\gamma \geq 2.4$ and refractory periods
correspondent to the memory effect, $T_r$. In addition, the
avalanche size distributions exhibit the power-law behavior in a
single scaling region in contrast to other networks. However, the
return time distributions displaying spatiotemporal complexity have
three characteristic time scaling regimes.
\end{abstract}

\end{frontmatter}

\newcommand{\be}{\begin{equation}}
\newcommand{\ee}{\end{equation}}

\section{Introduction}
 A human being is most complex organism in the nature. A human brain is extremely complicate in the organs of
human. That is the reason that it is exceedingly difficult to
understand brain through individual and classical mechanism. The
nervous tissue of the human brain contains many billions of neurons,
large complex cells that conduct nerve impulses from one part of the
body to another part. Scientists have investigated neural network in
the brain through various methods \cite{Sg01,ST02,ZC03,BR03,HB05}.
Recently, neural network was reported to be like the scale-free
network, which is characterized by displaying power law distribution
of the degree \cite{EC05}. An well known scale-free network is
Barab\'{a}i-Albert(BA) network characterized by evolving and
preferential attachment \cite{BA99,AB02}. Goh et al. introduce a
static scale-free network with no degree-degree correlation
\cite{GKK01}. The firing pattern of the neuron is similar to the
avalanche pattern of the self-organized criticality(SOC)
\cite{BTW87,BS93,JE98,Bak99}. The propagative size of neuron
functioning exhibits the power-law distribution. Thus, the
punctuated pattern and power-law behavior occur without fine-tuning
parameter. L. da Silva et al. offered a simple model for brain
functioning similar to the Bak-Sneppen(BS) model \cite{BS93} with
the memory effect on the lattice \cite{SPS98}. The same model is
investigated on small-world network \cite{WS98} by Lin and Chen
\cite{LC05}. We consider an avalanche dynamics of an idealized
neuron function on a uncorrelated random scale-free network.
 The structure and the dynamics of neurons describe as follows.
The anatomical unit of the nervous system is the neuron. The brain
possesses about $10^{10} - 10^{12}$ neurons. Neurons are quite
complex, but each of these is made up of dendrite, a cell body, and
an axon. A dendrite conducts signals toward the cell body. The cell
body is the part of a neuron that contains the nucleus and other
organelles. An axon conducts nerve impulses away from the cell body.
Dendrites and axons collectively are called neuron fibers. A nerve
impulse is the way a neuron transmits information. When an axon is
not conducting a nerve impulse, the resting potential indicates that
the inside of an axon is negative compared to the outside. In
contrast, if an axon is conducting a nerve impulse, an action
potential (i.e., electrochemical change) travels along a neuron. As
an axon is stimulated by an electric shock, threshold may be reached
for an action potential. A fiber can conduct a volley of nerve
impulses because only a small number of ions are exchanged with each
impulse. As soon as an impulse has passed by each successive portion
of a fiber, it undergoes a refractory period during which it is
unable to conduct an impulse. This ensures a one-way direction of
the impulse. During a refractory period, the sodium gate cannot yet
open. Every axon branches into $10^3 - 10^4$ fine terminal branches
called a synaptic bulb . Each bulb lies very close to the dendrite
(or the cell body) of another neuron. This region of close proximity
is called a synapse. At a synapse, the membrane of the first neuron
is called the presynaptic membrane, and the membrane of the next
neuron is called the postsynaptic membrane. Transmission of the
nerve impulse from one neuron to another takes place across a
synapse. In humans, synaptic vesicles release a chemical, known as a
neurotransmitter to receptors in the postsynaptic membrane increases
the chance of a nerve impulse (stimulation) or decreases the chance
of a nerve impulse (inhibition) in the next neuron \cite{SM98}. The
brain as a whole is a system capable of auto-regulations
\cite{SPS98}.

\section{ Model and Simulation Method }
 To make a modeling of this neuron dynamics, we make networks. The
networks is composed of uncorrelated random scale-free network.
Here, the uncorrelated random scale-free network is called by the
uncorrelated configuration model (UCM) \cite{CBS05}. We generate
$N$-cells by the static method. Let's make $N_k$-cells of the degree
$k$ satisfying the degree distribution $P(k)\sim k^{-\gamma}$.
Select two nodes randomly and connect them if they are not connected
before. We exclude duplicated connections and self-connection. The
network generated by the UCM is not only fully connected and have
but also no degree-degree correlation $\bar{D_{nn}}(k)$, defined as
the average degree of the nearest neighbors(NN) of the nodes with
degree k and clustering correlation $\bar{C}(k)$, mean by the
probability that a node of degree k form loop with two NN
\cite{CBS05}. The number of minimal degree is fixed at m=3 to
prevent alteration of dynamics from dangling node.
\begin{figure}
\includegraphics[width=10cm,height=12cm,angle=270,clip]{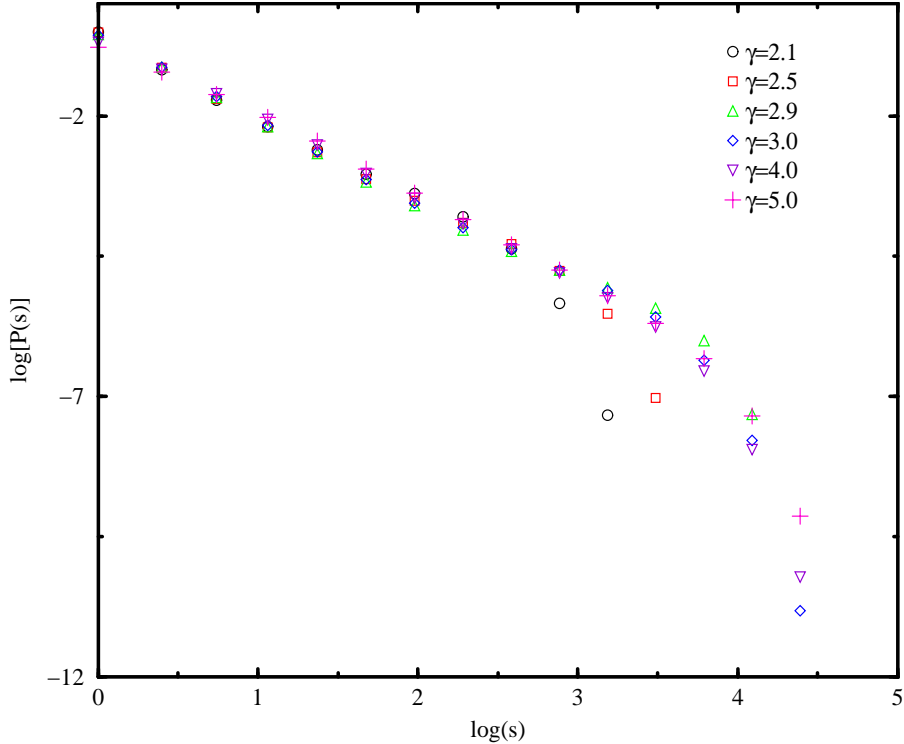}
\caption[0]{ The log-log plot of the probability distribution
function $P(s)$ of the $B_0(\gamma)$ avalanche size as a function of
the avalanche size $s$ at the critical fitness on UCM for $ \gamma =
2.1, 2.5, 2.9, 3.0, 4.0$, and $5.0$ with $N=10000$, $m=3$, and $T_r
= 1$. The critical threshold barriers where $B_0 = 0.039, 0.080,
0.121, 0.130, 0.212$, and $0.260$ respectively.} \label{fig1}
\end{figure}
 Now, we introduce
the evolution rule of neuron dynamics. Each node of the UCM
represents a neuron and a link between two nodes represents a
synapse. The uniform random numbers are distributed over each node.
This random number is called by a barrier, which is the possibility
of the firing of a neuron's spike. The lower barrier has a higher
potentialities to fire \cite{SPS98}. To simplify the firing is
occurred at an neuron with lowest barrier sequentially. And directly
connected neurons also is fired because they are enough stimulated
to fire. The selected neuron with the lowest barrier can not fire if
the elapsed time after the firing is less than the refractory period
$T_r$. In fact, neuron's spike transfer directly through their
synapse. However, the evolution of the signal is just considered to
spread out to all of nearest neighbors. If this process is iterated
the system reaches to critical stationary state, which all the
barriers are above the $B_c$ barrier so-called a critical threshold.
That is, neural system is self organized without well tuning
parameter to stationary state. And the brain functioning is occurred
abruptly underlying avalanche dynamics. The purpose of this research
is to investigate how the avalanche dynamics is changed according to
varying degree exponent $\gamma$ of networks with different
refractory periods $T_r$. For investigating spatiotemporal
correlation of avalanche dynamics, we examined first and all return
time distribution(RTDs).

\begin{figure}
\includegraphics[width=10cm,height=12cm,angle=270,clip]{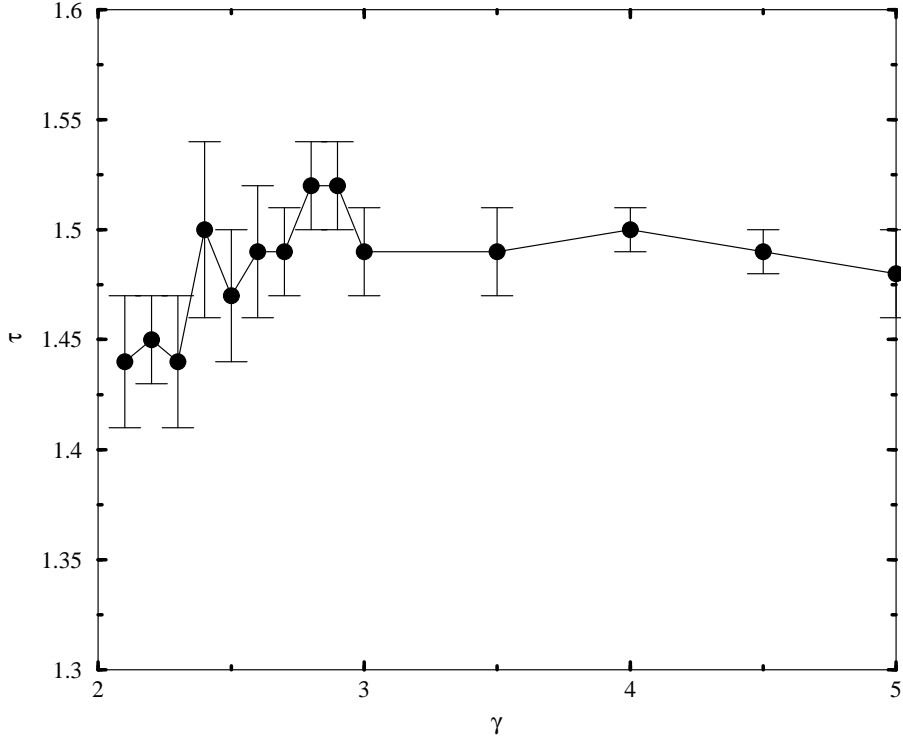}
\caption[0]{ Dependence of the avalanche size exponents as a
function of network degree exponents $\gamma$ with $N=10000$, $m=3$,
and $T_r = 1$.} \label{fig2}
\end{figure}
\begin{figure}
\includegraphics[width=10cm,height=12cm,angle=270,clip]{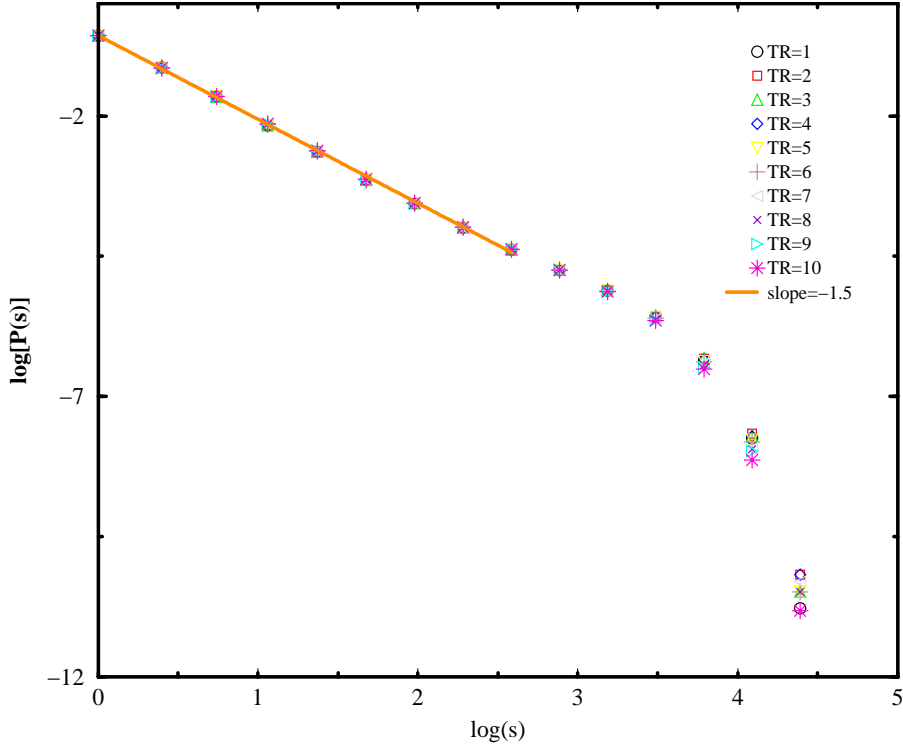}
\caption[0]{ The log-log plot of the probability distribution
function $P(s)$ of the $B_0(T_r)$ avalanche size as a function of
the avalanche size $s$ at the critical fitness on UCM from $T_r =1$
to $T_r=10$ with $N=10000$, $m=3$, and $\gamma = 3$. The critical
threshold barrier is $B_0 = 0.13$ for all $T_r$.} \label{fig3}
\end{figure}
\begin{figure}
\includegraphics[width=10cm,height=12cm,angle=270,clip]{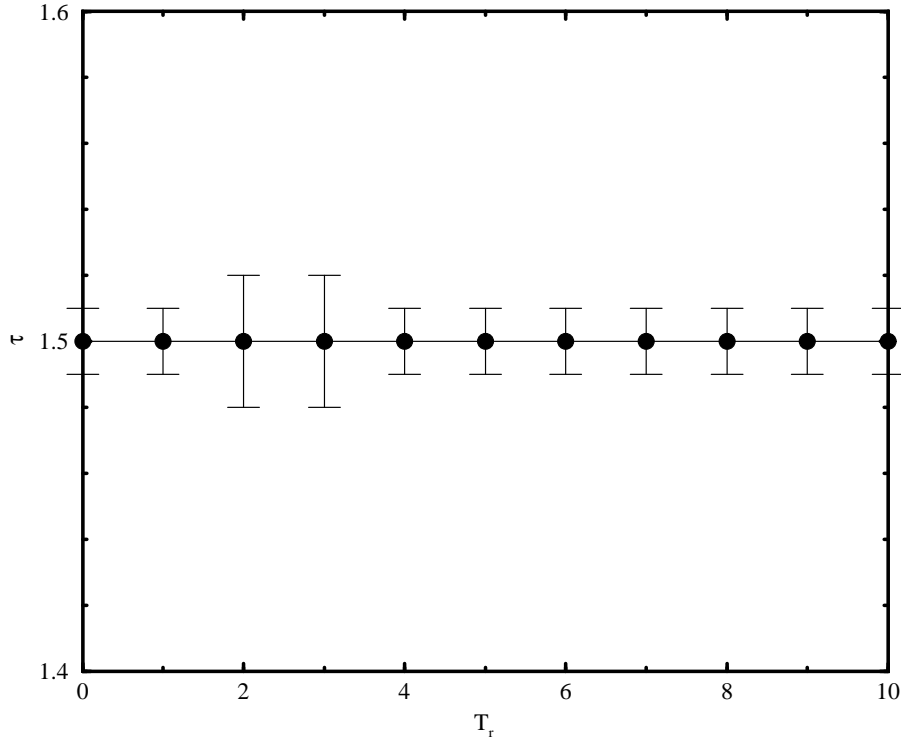}
\caption[0]{ Dependence of the avalanche size exponents as a
function of refractory periods $T_r$ with $N=10000$, $m=3$, and
$\gamma = 3$.} \label{fig4}
\end{figure}

\section{ Results }
 In stationary state, we consider the branching process of a avalanche dynamics to be unrestricted from network size
 corrections \cite{PMB96}. The avalanche is always started from the hub neuron
 and all of the neurons with $B_i > B_0$ , where $B_0$ is an auxiliary parameter, are treated as inactive
 neurons. A $B_0$ avalanche size $S$ is defined as the number of the firing less than $B_0$.
 As $B_0 \rightarrow B_c$, the avalanche size distribution
 follow a power-law behavior $P(S) \thicksim S^{-\tau}$ with an exponential cutoff.
Figure 1 shows avalanche size distribution for the $B_0(\gamma)$
avalanche with the degree exponents of the networks, $\gamma = 2.1,
2.5,  2.9,  3.0,  4.0$, and $5.0$ The avalanche size distribution
follows power-law behavior, $P(S) \sim S^{-\tau(\gamma)}$ extending
over more larger regime than on other scale-free networks such the
BA network \cite{BA99} and the static model introduced by Goh et al.
\cite{GKK01,LGKK04}. The more interesting thing is that the power
law behaviors of the avalanche size distribution do not exhibit the
crossover between two different scaling regimes. The avalanche size
distribution shows a short intermediate regime and follows
exponential decay at the cut-off regime. The absence of the two
regimes in the avalanche size distribution may be able to explain by
following two reasons. One of the reasons is that an average degree
$\langle k \rangle$ is not fixed with different degree exponent
$\gamma$ as compared with another static models by the Goh's
algorithm \cite{GKK01} and in addition is increased as $\gamma$ gets
smaller. Another reason is the absence of clustering correlations
 $\bar{C}(k)$ as well as degree-degree correlations
$\bar{D_{nn}}(k)$ \cite{CBS05}. Figure 2 presents the basic critical
exponents $\tau$, the so called avalanche size exponent from
different $\gamma$ ,where the increment is $0.1$ in $2 < \gamma < 3$
and $0.5$ in $ \gamma > 3$, with $T_r=1$. As we can observe, the
critical avalanche size exponent is the same as mean field result
i.e, $ \tau \simeq 1.5$ for $ \gamma \geq 2.4$. It is difficult to
compare the avalanche size critical exponent $\tau$ in the UCM with
the BA network because the avalanche size distribution on the BA
network shows different power-law behavior with two regimes
\cite{LK05,MGK05}. Occasionally, the critical thresholds are very
close value each other,  $f_c = 0.086\pm1$(on UCM) and $f_c =
0.089\pm2$(on BA) within error bar. Even though it is not universal
value, $f_c$ is seem to a criterion distinguishing different
avalanche dynamics of $ \gamma < 2.4$ from mean field result for
$\gamma \geq 2.4$.
 In the lattice \cite{SPS98} and small-world network \cite{LC05} with low rewiring probability $\phi =
 0.01$, the l\'{e}vy-flight exponents and the avalanche size exponents increase according to the increment of the refractory
 period $T_r$, respectively.
Figure 3 illustrates the probability distribution for the avalanche
size on UCM with different $T_r$. At the same time, in Fig.4, we
demonstrate the dependence of the exponent $\tau$  from 1 to 10 with
$B_0 = 0.13$. All $\tau$ are not changed according to varying $T_r$
unlike the results on lattice \cite{SPS98} or small-world network
\cite{LC05}. In the lattice, the avalanche dynamics of the firing is
propagated further far away from first update neuron because the
firing is rejected during $S < T_r$ as refractory period $T_r$
increase. But in scale-free network, the active neuron returns soon
to the hub and the direct linked node in stationary state because a
hub neuron has many nearest neighbor's neurons. Accordingly, the
firing neuron is not evolved far from hub quickly with refractory
period. For that reason, the memory effects of refractory time
vanish as the existence of hub is growing larger. In case of
small-world network, the rewiring probability approaches a threshold
to eliminate memory effect, the exponent $\tau$ also follows the
mean-field value for different $T_r$.

\begin{figure}
\includegraphics[width=10cm,height=12cm,angle=270,clip]{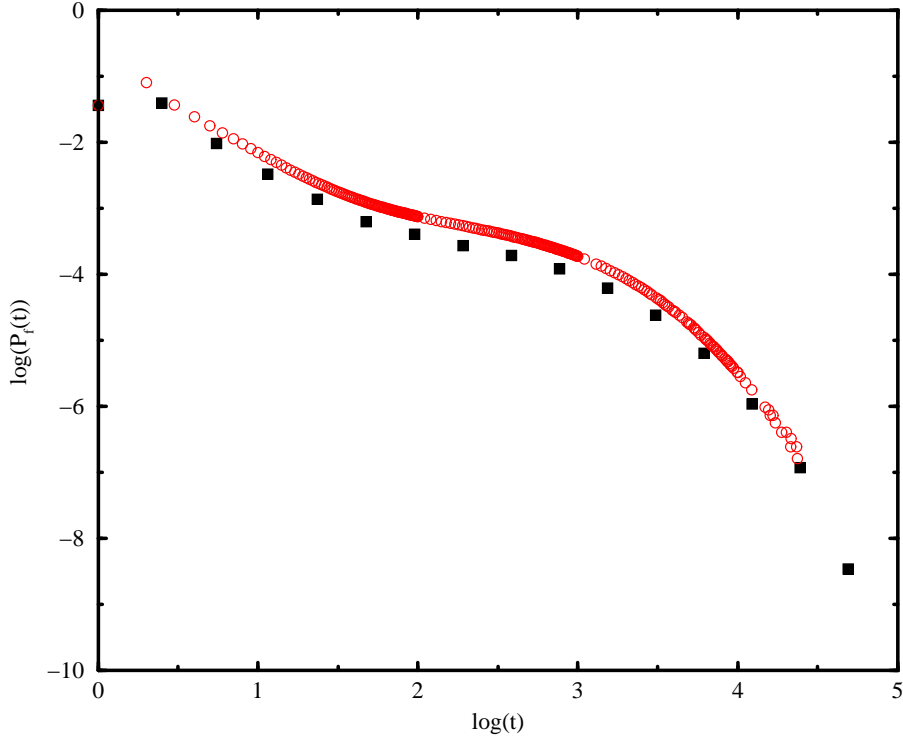}
\caption[0]{ The log-log plot of the probability distribution
function $P(s)$ of the first return time  as a function of the time
$t$ at the critical fitness on UCM with $N=10000$, $m=3$ at $T_r
=1$, and $\gamma = 3$ The solid symbol presents the histogram, using
the exponential bin plot.} \label{fig5}
\end{figure}
\begin{figure}
\includegraphics[width=10cm,height=12cm,angle=270,clip]{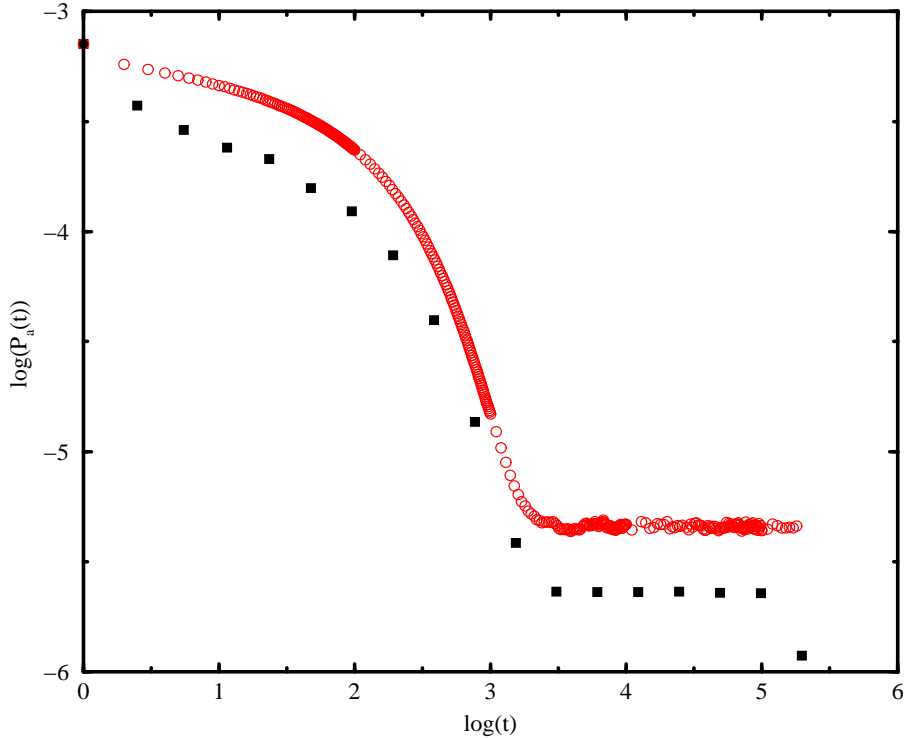}
\caption[0]{ The log-log plot of the probability distribution
function $P(s)$ of the all return time as a function of the time $t$
at the critical fitness on UCM with $N=10000$, $m=3$ at $T_r =1$,
and $\gamma = 3$ The solid symbol presents the histogram, using the
exponential bin plot.} \label{fig6}
\end{figure}

  The probability distribution of first and all return time is
 valuable quantities for investigating the spatiotemporal
 correlation and punctuated pattern \cite{BS93,JE98,Bak99,PMB96}. The definition of first return
 time with the size t is a separating intervals which
 activated subsequently from a given active neuron.
In Fig.5, we present first return time distribution(FRTd) for
$\gamma=3$, $T_r=1$. The first return time distribution do not
satisfy power-law behavior contrary to the lattice case. The
power-law behavior of the early return time region is mostly
affected by the dynamics of a hub node as the hub and nearest
neighbors frequently fire in the ratio of $k_{hub} + 1$ ,which
$k_{hub}$ is the degree of the hub and $\langle k_{hub} \rangle =
45$ in Fig.5, that is, the hub has a probability of the firing as
$1/(k_{hub} + 1)$. An interval of power-law regime in early return
time increases with the degree of the hub. The intermediate return
time distribution become almost constant for all nodes as $N
\rightarrow \infty$ because each node has a same probability to be
active again since UCM do not have the degree correlations. Finally,
long return time has a long exponential decay regime with cut-off
for $t \rightarrow L\xi$ by the finite size effect of the dangling
node because the diameter of scale-free network is very small. All
return times with the size t is the elapsed time steps to time $t$
regardless of the intermediate firing since a given neuron fired at
time $t_0$. In Fig. 6, we plot all return time distribution for
$\gamma=3$, and $T_r=1$. The all return time distribution(ARTd) also
is divided up three characteristic time scaling regimes.
Furthermore, the slope of each regime is small in early return time
regime like the lattice model although the general scaling relation
is not satisfied.

\section{Conclusion}
 We have studied the simple model of the neuron function
in the brain on the uncorrelated scale-free network. Our model show
the avalanche dynamics with different degree exponent $\gamma$ of
the networks and different refractory periods $T_r$. The avalanche
size exponent, $\tau$ is not changed according to varying $\gamma$,
for $\gamma \geq 2.4$ as well as $T_r$. We measured first and all
return time distribution(RTDs). The RTDs do not follow power-law
behaviors consistently and show the three characteristic regions. In
future work, the simple model for the SOC system on the scale-free
network that memory effects contribute would be inquired and a kind
of scale-free network more neural system-like should be
investigated.

\section*{Acknowledgments}
 This work was supported by Inha University Research Foundation
 Grant.

\newcommand{\jpa}{J. Phys. A}
\newcommand{\jkps}{J. Kor. Phys. Soc.}

\end{document}